\documentclass[prl, twocolumn, showpacs, amsmath, amssymb]{revtex4}
\usepackage{dcolumn}
\usepackage{bm}
\usepackage{graphicx}
\usepackage{color}
\DeclareGraphicsExtensions{. jpg, . pdf, . mps, . png, . eps, . ps, . EPS}

\begin{document}
\def\be{\begin{equation}}
\def\ee{\end{equation}}
\def\bc{\begin{center}} 
\def\ec{\end{center}}
\def\bea{\begin{eqnarray}}
\def\eea{\end{eqnarray}}
\newcommand{\avg}[1]{\langle{#1}\rangle}
\newcommand{\Avg}[1]{\left\langle{#1}\right\rangle}

\title{Entropy measures for networks: Toward  an information theory of complex topologies}
\author{Kartik Anand$^1$ and Ginestra Bianconi$^2$}
\affiliation{ $^1$The Abdus Salam International Center for Theoretical Physics, 
 Strada Costiera 11, 34014 Trieste, Italy\\
$^2$Departement of Physics, Northeastern
University,  Boston, Massachusetts 02115 USA }
\begin{abstract}
The quantification of the complexity of networks is, today, a fundamental problem in the physics of complex systems. A possible roadmap to solve the problem is via extending key concepts of information theory to networks. In this paper we propose how to define the Shannon entropy of a network ensemble and how it relates to the Gibbs and von Neumann entropies of network ensembles.  The quantities we introduce here will play a crucial role for the formulation of null models of networks through maximum-entropy arguments and will contribute to inference problems emerging in the field of complex networks. 
\end{abstract}

\maketitle 

Complex networks \cite{Evolution, Latora, Doro_c,Santo} are found to characterize the underlying structure of many biological, social and technological systems. Following ten years of active research in the field of complex networks,  the state of the art includes, a deep
understanding of their evolution \cite{Evolution}, an unveiling of the
rich interplay between network topology and dynamics \cite{Doro_c} and
a description of networks through structural characteristics
\cite{Latora,Santo}. Nevertheless, we still lack the means to
quantify, {\it how complex is a complex network}. In order to answer
this question we need a new theory of information of complex
networks. This new theory will contribute to solving many challenging
inference problems in the field \cite{Gfeller,Features,Santo}. By providing an evaluation of the information encoded in complex networks, this will resolve one of the outstanding problems in the statistical mechanics of complex systems. 

In information theory \cite{TCover} entropy measures play a key role. In fact, it is well known that the Shannon entropy and the von Neumann
entropy are related to the information present in classical and
quantum systems, respectively. Moreover, the afore mentioned measures
also have statistical mechanics interpretations.  Traditionally, in statistical
  mechanics, for configurations drawn from canonical ensembles, the Shannon entropy corresponds to the entropy for classical systems, while the von Neumann entropy provides the statistical description of quantum systems.

In the context of complex networks a number of different entropy
measures have been introduced \cite{Burda_en, Gfeller,entropy,
  Latora_rate, Ldev, BGS, Rovelli}. In Ref. \cite{entropy} the Gibbs entropy per node, in a network of $N$ nodes, denoted $\Sigma$, was introduced for microcanonical network ensembles following a statistical mechanics paradigm. Microcanonical network ensembles are defined as those
networks that satisfy a given set of constraints. Examples of some popular constraints include, fixed number of links per node, given degree sequence and community structure. The Gibbs entropy of these ensembles is given by 
\be
\Sigma=\frac{1}{N}\log {\cal N}\,,
\label{uno}
\ee 
where ${\cal N}$ indicates the cardinality of the ensemble, i.e., the total number of networks in the ensemble. As demonstrated further in \cite{entropy} the statistical mechanics formalism enables us to develop canonical network ensembles where the structural constraints under consideration are satisfied, {\it on average}. In classical statistical mechanics the microcanonical ensemble is formed by configurations having constant energy $E$, while the canonical ensemble is formed by configurations having constant average energy $\avg{E}$. By analogy, in the theory of random graphs the $G(N,L)$ graph ensemble is formed by networks of $N$ nodes with a
constant total number of links $L$. In the conjugated-canonical $G(N, p)$ ensemble, however, the total number of links is Poisson distributed
with average $\avg{L}=p(N-1)$. This construction of microcanonical and conjugate-canonical ensemble can be further generalized \cite{entropy} to network ensembles with more elaborate sets of constraints. For example we can define microcanonical network
ensembles with given degree sequence $\{\kappa_i\}$ and canonical network ensembles (based on hidden variables \cite{hv, Boguna_hv}) in which each node $i$ has $k_i$ links, which is Poisson distributed with average $\avg{k_i}=\kappa_i$. 

In this letter we show {  for this new
statistical mechanics framework of networks}, first, that the entropy
of canonical network ensembles is related to the Shannon Entropy and
second, that canonical network ensembles satisfy a principle of
maximal Shannon entropy. Moreover we will study to what extent
canonical and microcanonical network ensembles are equivalent. Finally
we will discuss the relation between the Shannon entropy of a
canonical network ensemble, $S$, and the recent definition of von
Neumann entropy of networks, $S_{VN}$, recently introduced in
Ref.\cite{BGS} of interest in the field of quantum gravity \cite{Rovelli}.

{\it Gibbs entropy of a microcanonical network ensemble. }
Microcanonical network ensemble are formed by network satisfying a given number of constraints. Following the lines of reasoning provided in \cite{entropy}, on specifying the full set of constraints and number of nodes $N$ in the networks, one may introduce a partition function $Z$ for the ensemble. This partition function counts the number of networks, defined by their adjacency matrices $\{a_{ij}\}$, that simultaneously satisfy all the constraints under consideration. The adjacency matrix describes an undirected network, i.e., $a_{ij}=a_{ji}$, where each element takes some positive integer values, $a_{ij}\in\alpha$, where $\alpha\subset\mathbb{N}$, that indicates the weight of a link between nodes $i$ and $j$. For simple (connectivity) networks we take $a_{ij}\in\{0,1\}$ while for weighted networks $a_{ij}\in\mathbb{N}$. Thus, we have
\be
Z=\sum_{\{a_{ij}\}}\prod_k\delta(\mbox{constraint}_k(\{a_{ij}\}))e^{-\sum_{i<j}\sum_{\alpha}h_{ij}(\alpha)\delta_{a_{ij}, \alpha}}\,,
\ee
where the fields $h_{ij}(\alpha)$ play the usual role of auxiliary fields in statistical mechanics. Finally the Gibbs entropy $\Sigma$, defined by Eq. $(\ref{uno})$, and the probability $\pi_{ij}(\alpha)$ of having a link between nodes $i$ and $j$, with weight $\alpha$, are given by 
\bea
N\Sigma&=&\left. \log Z\right|_{h_{ij}(\alpha)=0 \ \forall (i, j,\alpha)}\,,\nonumber\\
\pi_{ij}(\alpha)&=&\frac{\partial \log Z}{\partial h_{ij}(\alpha)}\,.
\label{micro}
\eea
{\it Entropy of a canonical network ensembles. }
The canonical network ensemble can be built starting from the marginal distribution $\pi_{ij}(\alpha)$, given by Eq. $(\ref{micro})$. For a network of $N$ nodes, for each pair of nodes, $(i,j)$, one draws a link of weight $\alpha$ with probability $\pi_{ij}(\alpha)$. 
The probability $\Pi$ of the canonical undirected network ensemble, defined by its adjacency matrix $\{a_{ij}\}$, is therefore given by
\be
{\Pi}=\prod_{i<j}\pi_{ij}(a_{ij})\,,
\ee
for which the log-likelihood function is given by
\be
{\cal{L}}=-\sum_{i<j}\log \pi_{ij}(a_{ij})\,.
\ee
The entropy of a canonical ensemble is the logarithm of the number of typical networks in the ensembles and is given by 
\be
S=\avg{{\cal{L}}}_{\Pi}=-\sum_{i<j}\sum_{\alpha}\pi_{ij}(\alpha)\log
\pi_{ij}(\alpha)\,,
\label{Shannon_w}
\ee
which takes exactly the form of a Shannon entropy. We will therefore call this quantity the Shannon entropy of a network ensemble. 
In particular, for the case of a simple undirected network, where $\alpha\in\{0,1\}$, we have 
\be
S=-\sum_{i<j}p_{ij}\log p_{ij}-\sum_{i<j}(1-p_{ij})\log(1-p_{ij})\,,
\label{Shannon}
\ee
where $p_{ij}=\pi_{ij}(1)$ is the probability of having a link between nodes $i$ and $j$. 

Maximizing the Shannon entropy of the network subjected to different
types of constraints gives rise to maximum-entropy ensembles and
generalizing the maximum-likelihood arguments of \cite{Garlaschelli_ML}. In the following we will consider few examples, of such constraints for the cases of simple undirected networks. 

Fixing the total number of expected links, $\sum_{ij}p_{ij}=L$, the
maximum-entropy ensemble is $G(N,\{p_{ij}\})$, with
$p_{ij}=p=L/(N(N-1)/2)$. Alternatively, if we constrain the expected
degree of each node $i$, i.e., $\kappa_i=\sum_{j}p_{ij}$, the
probabilities in the maximum-entropy ensemble take the form
$p_{ij}=\theta_i\theta_j/(1+\theta_i\theta_j)$ where $\theta_i$ are
hidden-variables fixed by the constraints. This
ensemble is the canonical conjugated to the microcanonical ensemble of
networks with fixed degree sequence $\{\kappa_i\}$.  In table I we
generalize this construction and report the form of maximum-entropy
network ensembles satisfying a different sets of constraints. We leave
to the reader the construction of  maximum-entropy weighted network ensembles related to the
canonical ensembles discussed in
Refs. \cite{entropy,Garlaschelli_w}. { The marginal probability for
  the microcanonical and conjugated canonical ensembles are equal by
  definition, but in order to prove  the equivalence between the two
  ensembles  also the entropy per node $\Sigma$ and $S/N$ must be equal in the thermodynamic limit.}
 
\begin{table*}
\begin{ruledtabular}
\begin{tabular}{|c|c|c|c|}
 Ensembles &Probabilities $p_{ij}/(1-p_{ij})$
 &\multicolumn{2}{c|}{Conditions} \\\hline 
$\begin{array}{c}\mbox{Given expected}\\ \mbox{ number of links }
 L\end{array}$ & $p/(1-p)$& $pN(N-1)/2=L$ & \\\hline$\begin{array}{c}\mbox{Given expected}\\\mbox{ community structure} \{A_{q, q'}\}\end{array}$&
${W(q_i, q_j)}$ & &
$\begin{array}{c}\left. A(q, q')\right|_{q\neq
 q'}=\sum_{ij}p_{ij}\delta_{q_i, q}\delta_{q_j, q'}
 \\ A(q, q)=\sum_{i<j}p_{ij}\delta_{q_i, q}\delta_{q_j, q}\end{array}$
\\\hline $\begin{array}{c}\mbox{Given expected}\\\mbox{ degree
 sequence }\{\kappa_i\} \end{array}$&$\theta_i\theta_j$&$\kappa_i=\sum_{j}p_{ij}$ & \\\hline
$\begin{array}{c}\mbox{Given expected}\\\mbox{ degree sequence } \{\kappa_i\}\\\mbox{ community structure } \{A(q, q')\}\end{array}$&${\theta_i
 \theta_j W(q_i, q_j)}$& $\kappa_i=\sum_{j}p_{ij}$&
$\begin{array}{c}\left. A(q, q')\right|_{q\neq
 q'}=\sum_{ij}p_{ij}\delta_{q_i, q}\delta_{q_j, q'}
 \\A(q, q)=\sum_{i<j}p_{ij}\delta_{q_i, q}\delta_{q_j, q}\end{array}$ \\\hline $\begin{array}{c}\mbox{Given expected }\\ \mbox{degree
 sequence } \{\kappa_i\} \\ \mbox{and number of link at distance } d, B(d)\end{array}$& $\theta_i
\theta_jW(d_{ij}) $& $\kappa_i=\sum_{j}p_{ij}$ & $B(d)=\sum_{ij}p_{ij}\delta_{d_{ij}, d}$\\\hline
$\begin{array}{c}\mbox{Given expected}\\\mbox{ degree sequence }
 \{\kappa_i\} \\ \mbox{and number of triangles }\\\mbox{ for each node } \{T_i\} \end{array}$&
${\theta_i\theta_j e^{f_{ij}(\alpha_i+\alpha_j)+g_{ij}}}$ &
$\begin{array}{c} \kappa_i=\sum_{j}p_{ij}\\
 T_i=\sum_{jk}p_{ij}p_{jk}p_{ki}\end{array}$
&$\begin{array}{c} f_{ij}=\sum_{k}p_{ik}p_{kj}\\
 g_{ij}=\sum_kp_{ik}\alpha_kp_{kj} \end{array}$
\end{tabular}
\end{ruledtabular}
\caption{Maximum-entropy networks ensembles with given set of
 constraints. The community of each node is associated with a Potts
 variable $q_i$. The distance of the nodes is binned and indicated by
 a discrete variable $d_{ij}=d$. The hidden variables of each
 ensembles $\{\theta_i\}$, 
 $W(q, q')$, $W(d)$, $\{\alpha_i\}$, $\{f_{ij}, g_{ij}\}$, are fixed by
 respective conditions specified in the table. }
\label{table}
\end{table*}

{\it Comparison between the entropies of the $G(N,L)$ and the $G(N,p)$ ensembles. }
We study first the relation between the Gibbs entropy $\Sigma$ and the Shannon entropy per node for random graphs, defined for the $G(N, L)$ and $G(N, p)$ ensembles, respectively. The Gibbs entropy in the $G(N, L)$ ensemble is given by \cite{Burda_en}
\be
N\Sigma=\log \left(\begin{array}{c} \frac{N(N-1)}{2} \\
 L\end{array}\right)\,.
\label{SGNL}
\ee
As mentioned earlier, the corresponding probability of each link in the conjugate $G(N,p)$ ensemble is given by $p_{ij}=p=2L/(N(N-1))$. Inserting this probability in the definition of the Shannon entropy,  Eq. $(\ref{Shannon})$, we get 
\bea
\Sigma&=&S/N+\frac{1}{2N}\left[\log\left(\frac{N(N-1)}{2L}\right)-\log\left(\frac{N(N-1)}{2}-L\right)\right]\,.\nonumber
\eea
Therefore the Gibbs entropy $\Sigma$ and the Shannon entropy per node
$S/N$ of random graphs are equal in the thermodynamic limit $N\rightarrow \infty$. 

{\it Comparisons between the network ensembles with given degree sequence and structural cutoff. }
The microcanonical ensemble of networks with given degree sequence $\{\kappa_i\}$ has been fully characterized in \cite{entropy}. 
For simplicity, we consider the Gibbs entropy per node $\Sigma$ in the case where the maximal connectivity of the nodes satisfy a structural cutoff, i.e., $k_{max}<\sqrt{\avg{\kappa}N}$. In this limit the statistical mechanics treatment gives the Bender formula \cite{Enumeration} and the Gibbs entropy per node $\Sigma$ is given by 
\be
N\Sigma=\log[(2L-1)!!]-\sum_{i}{\log(\kappa_i!)}-\frac{1}{4}\left(\frac{\sum_i\kappa_i^2}{\sum_i
 \kappa_i}\right)^2\,.
\ee
In the conjugate-canonical ensemble, the probability of having a link is given by $p_{ij}=\frac{\kappa_i \kappa_j}{\avg{\kappa}N}$. Inserting this expression into Eq. $(\ref{Shannon})$ we get for the Shannon entropy of the ensemble
\be
\Sigma=S/N-\sum_i [\log(\kappa_i!/(\kappa_i^{\kappa_i}e^{-\kappa_i}))]+{\cal O}{(\log(N)/{N})}
\ee
We observe that the entropy per node $\Sigma$ and the Shannon entropy per node $S/N$ of the canonical conjugated network ensemble are not equal in the thermodynamic limit. This implies, for example, that the entropy per node of regular networks is smaller
than that of a Poisson network with same average degree. In particular, suppose we take, for regular networks, $\kappa_i=c$ and for the conjugated-canonical Poisson network, $k_i$ to be a Poisson distributed random variable with a mean $\avg{k_i}=\kappa_i=c$. 
The entropy of regular networks $\Sigma_R$ and the entropy of Poisson networks $S_{ER}$ are related by the expression
\bea
\Sigma_R=S_{ER}/N-\log\frac{c!}{c^ce^{-c}}\simeq S_{ER}/N-\frac{1}{2}\log(c)\,,\eea
where in the last expression we have taken the Stirling approximation valid for large $c$. 

The non-equivalence of $\Sigma$ and $S/N$ in the thermodynamic limit
can be also checked for network ensembles  satisfying further
constraints  as for example the networks ensembles with given degree
sequence and network community structure, and network ensembles with given degree sequence and given spatial dependence of the networks on the distance between the nodes. 
In general it is possible to  demonstrate that as soon as we consider ensembles of networks with an extensive number of constraints the Gibbs entropy per node $\Sigma$ and the Shannon entropy per node $S/N$ are non-equal in the thermodynamic limit. 

{\it The von Neumann entropy of a network ensemble. }
In \cite{BGS} the authors  have shown that is possible to define a von Neumann entropy of a network. This entropy is constructed from a density matrix $\rho$ associated with the network. The density matrix must be a positive semi-definite matrix with unitary
trace. In order to construct a density matrix from a network,in
\cite{BGS} it is proposed to consider the matrix $\rho={L}/{\sum_{ij}a_{ij}}$, where $L$ is the Laplacian matrix of the network, with $L_{ij}=\sum_{r} a_{ir}\delta_{i, j}-a_{ij}$. The spectrum \cite{Sam} of the Laplacian matrix is important for the  stability of $O(n)$ models, synchronization properties of networks and determining the scaling of the return times of random walk on the network \cite{Burioni}.  Given $\rho$ as specified above, we can calculate the average von Neumann entropy of an ensemble as
\be
S_{VN}=-\avg{\mbox{Tr} \rho\log(\rho)}_{\Pi}\,.
\ee
The von Neumann entropy is therefore related to the spectra of the Laplacian. The theoretical evaluation of the self-averaging spectra of the
Laplacian of complex networks ensemble is a very challenging topic that has attracted recent interest in the statistical mechanics community \cite{Sam}. Here we numerically explore how the von Neumann entropy $S_{VN}$ is related to the Shannon entropy of canonical ensembles. 

\begin{figure}
\includegraphics[width=80mm, height=60mm]{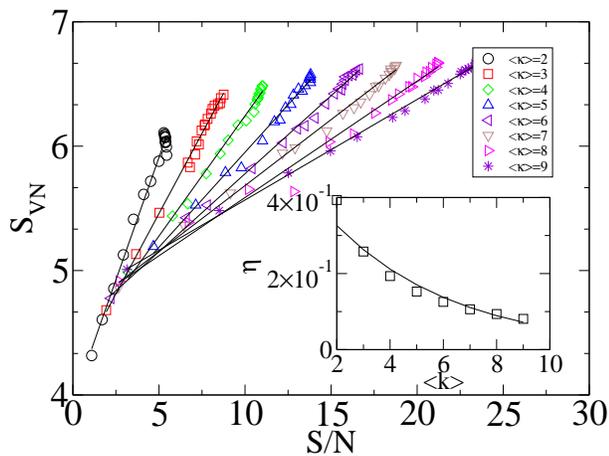}
\caption{(Color online)The von Neumann entropy $S_{VN}$ versus the Shannon entropy
 per node $S/N$ calculated for ensembles of scale-free networks with different
 expected average degree $\avg{\kappa}$. The points are calculated by 
 averaging over $20$ networks in the ensemble of networks with
 $N=1000$ nodes and different 
 power-law exponents $\gamma$ of the distribution of the expected
 degrees $P(\kappa)\propto \kappa^{-\gamma}$. The inset report the slope
 $\eta$ defined in
 $ [\ref{eta}]$ as a function of $\avg{\kappa}$ and the exponential
 fit indicated as a  solid line. }
\label{fig2. fig}
\end{figure}

For $G(N, p)$ networks the average von Neumann entropy, $S_{VN}$, is
an increasing function of the average connectivity, $pN$, while the
Shannon entropy per node, $S/N$, has the typical bell-shape form given
by Eq. $(\ref{SGNL})$, in the limit of large $N$. Therefore, for the
$G(N, p)$ random graphs ensembles,  the relation between $S_{VN}$
and $S$ is non monotonic, when we vary the average connectity $p(N-1)$. It is instructive to study the relation of the Shannon entropy of a network ensemble and its average von Neumann entropy in networks with the same average degree. In \cite{entropy} it has been shown that networks with power-law degree distribution $P(k)\propto k^{-\gamma}$ and constant average degree $\avg{k}$ have a Gibbs entropy per node $\Sigma$ which is an increasing function of the power-law exponent $\gamma$. Similarly the Shannon entropy per node $S/N$ of canonical network ensembles with fixed expected
degree $\kappa_i$, where $P(\kappa)\propto \kappa^{-\gamma}$ and fixed $\avg{\kappa}$ is increasing with the power-law exponent $\gamma$.   Therefore changing the power-law exponent $\gamma$ is a way to modulate $S$ by leaving the average degree constant. 
In figure $\ref{fig2. fig}$ we report the von Neumann entropy $S_{VN}$ vs. the Shannon entropy per node $S/N$ in canonical power-law network ensembles with constant $\avg{\kappa}$ and variable value of the $\gamma$ exponent.  We find that the two entropies are linearly related
\be
S_{VN}=\eta S/N+\beta\,,
\label{eta}
\ee
where $\eta$ decays exponentially as a function of $\avg{\kappa}$ for small values of $\avg{\kappa}\ll N$. Therefore for scale free networks the von Neumann entropy is linearly related to the Shannon entropy of the canonical ensembles measuring the number of typical networks in the ensemble.

{\it Conclusions }
In this Letter we have explored the connection between different
definition of entropy of network ensembles. Interesting we have found
that the Gibbs entropy per node $\Sigma$ is equal to the Shannon
entropy per node $S/N$ in the thermodynamic limit for random
graphs. However, when we consider networks with and extensive number
of constraints (as for example a given degree distribution) the Gibbs
entropy per node $\Sigma$ and the Shannon entropy per node $S/N$
differ by ${\cal O}(1)$ terms. Moreover we have related the Shannon
entropy with the recently introduced von Neumann entropy of
networks. Interestingly we found that for scale free networks with
constant average degree $S_{VN}$ and $S/N$ are linearly related. We
believe that all the entropies of the network ensembles, $S$ and
$S_{VN}$ as well as $\Sigma$ \cite{Features} will play a crucial role
for the quantification of the complexity and in inference problems in
networks. 

 G. B.  acknowledge stimulating discussions with  J. Baranyi, A.C.C. Coolen and A. N. Samukhin.


\begin{thebibliography}{99}

\bibitem{Evolution}
R. Albert and A. -L. Barab\'asi, Rev. Mod. Phys. {\bf 74}, 47 (2002);
S. N. Dorogovtsev and J. F. F. Mendes, {\it Evolution of networks: From
 Biological Nets to the Internet and the WWW}
(Oxford University Press, Oxford, 2003); M. E. J. Newman, SIAM Review
{\bf 45}, 167 (2003).
\bibitem{Latora} S. Boccaletti, V. Latora, Y. Moreno, M. Chavez and
D. U. Hwang, Phys. Rep. {\bf 424}, 175 (2006).

\bibitem{Doro_c} S. N. Dorogovtsev, A. Goltsev and
J. F. F. Mendes, Rev. Mod. Phys. {\bf 80}, 61 (2008);
A. Barrat, M. Barth\'elemy and A. Vespignani, 
{\it Dynamics Processes on Complex Networks}
(Cambridge University Press, Cambridge, 2008).
\bibitem{Santo} S. Fortunato preprint
arXiv:0906.0612 (2009). 


\bibitem{Gfeller}
D. Gfeller, J.-C. Chappelier and P. De Los Rios, Phys. Rev. E {\bf
  72}, 056135 (2005).
\bibitem{Features}
G. Bianconi, P. Pin and M. Marsili, PNAS, {\bf 106},11433 (2009). 
\bibitem{TCover} T. Cover and J. Thomas, {\it Elements of Information Theory}, 
Wiley-Interscience (1991). 
\bibitem{Burda_en} L. Bogacz, Z. Burda and B. Waclaw, Physica A {\bf
 366}, 587 (2006). 
\bibitem{entropy} G. Bianconi, { Europhys. Lett. } {\bf 81}, 28005
(2008); G. Bianconi, Phys. Rev. E {\bf 79}, 036114 (2009). 
\bibitem{Ldev}
G. Bianconi, A. C. C. Coolen and C. J. Perez Vicente, Phys. Rev. E {\bf
78}, 016114 (2008). 
\bibitem{Latora_rate} J. Gomez-Gardenes and V. Latora, Phys. Rev. E {\bf
 78}, 065102 (2008). 
\bibitem{BGS}
S. L. Braustein, S. Gosh, S. Severini,Ann. of Combinatorics {\bf 10} 291, (2006);
F. Passerini and S. Severini, preprint arXiv:0812.2597 (2008). 
\bibitem{Rovelli}
C. Rovelli and F. Vidotto, preprint arXiv:0905.2983 (2009). 
\bibitem{hv}G. Caldarelli, A. Capocci, P. De Los Rios and
M. A. Mu\~noz Phys. Rev. Lett. {\bf 89}, 258702 (2002); J. Park and M. E. J. Newman Phys. Rev. E {\bf 70}, 066146 (2004).
\bibitem{Boguna_hv} M. Bogu\~n\'a and R. Pastor-Satorras Phys. Rev. E
{\bf 68}, 036112 (2003). 
\bibitem{Garlaschelli_ML}
D. Garlaschelli and M. I. Loffredo, Phys. Rev. E {\bf 78}, 015101(R)
(2008). 
\bibitem{Garlaschelli_w}
D. Garlaschelli and M. I. Loffredo, Phys. Rev. Lett. {\bf 102}, 038701 (2009). 
\bibitem{Enumeration} E. Bender and E. Rodney Canfield, 
J. Combin. Theory Ser. A {\bf 24}, 296 (1978); B. D. McKay Ars Combin. {\bf 19A}, 15 (1985). 
\bibitem{Sam} A. N. Samukhin, S. N. Dorogovtsev and J. F. F. Mendes, Phys. Rev. E
{\bf 77}, 036115 (2008); R. K\"uhn, J. Phys. A: Math. Theor. {\bf 41}, 295002 (2008); G. Bianconi preprint arXiv:0804.17441 (2008). 
\bibitem{Burioni} R. Burioni, D. Cassi and A. Vezzani, Phys. Rev E {\bf 60}, 1500 (1999);
M. Barahona and L. M. Pecora, Phys. Rev. Lett. {\bf 89}, 054101 (2002). 

\end{thebibliography}
\end{document}